\documentclass[12pt,preprint]{aastex}

\shorttitle{NGC~5548 in a Low-Luminosity State}
\shortauthors{Bentz, et al.}

\received{}
\accepted{}

\begin{document}

\title{NGC~5548 in a Low-Luminosity State: Implications for the Broad-Line Region}

\author{ Misty~C.~Bentz\altaffilmark{1}, 
         Kelly~D.~Denney\altaffilmark{1},
         Edward~M.~Cackett\altaffilmark{2,3},
         Matthias~Dietrich\altaffilmark{1},
         Jeffrey~K.~J.~Fogel\altaffilmark{3},
         Himel~Ghosh\altaffilmark{1}, 
	 Keith~D.~Horne\altaffilmark{2},
         Charles~Kuehn\altaffilmark{1,4},
         Takeo~Minezaki\altaffilmark{5},
         Christopher~A.~Onken\altaffilmark{1,6},
         Bradley~M.~Peterson\altaffilmark{1},
         Richard~W.~Pogge\altaffilmark{1},
         Vladimir~I.~Pronik\altaffilmark{7,8},
         Douglas~O.~Richstone\altaffilmark{3},
         Sergey~G.~Sergeev\altaffilmark{7,8},
         Marianne~Vestergaard\altaffilmark{9},
         Matthew~G.~Walker\altaffilmark{3}, and
         Yuzuru~Yoshii\altaffilmark{5,10}}

\altaffiltext{1}{Department of Astronomy, 
		The Ohio State University, 
		140 West 18th Avenue, 
		Columbus, OH 43210; 
		bentz, denney, dietrich, ghosh, peterson, 
                pogge@astronomy.ohio-state.edu}

\altaffiltext{2}{School of Physics and Astronomy, 
		 University of St.~Andrews,
		 Fife, KY16 9SS, Scotland, UK;
		 emc14, kdh1@st-and.ac.uk}

\altaffiltext{3}{Department of Astronomy,
		 University of Michigan,
		 Ann Arbor, MI 48109-1090;
		 fogel, dor, mgwalker@umich.edu}

\altaffiltext{4}{Present address:
		 Physics and Astronomy Department,
		 3270 Biomedical Physical Sciences Building,
		 Michigan State University,
		 East Lansing, MI 48824;
		 kuehncha@msu.edu}

\altaffiltext{5}{Institute of Astronomy, 
		 School of Science, 
		 University of Tokyo,
	 	 2-21-1 Osawa, Mitaka, 
		 Tokyo 181-0015, Japan;
		 minezaki, yoshii@mtk.ioa.s.u-tokyo.ac.jp}

\altaffiltext{6}{Present address:
		National Research Council Canada, 
		Herzberg Institute of Astrophysics,
		5071 West Saanich Road,
		Victoria, BC  V9E 2E7, Canada;
		christopher.onken@nrc-cnrc.gc.ca }

\altaffiltext{7}{Crimean Astrophysical Observatory,
		 p/o Nauchny, 98409 Crimea, Ukraine;
		 sergeev, vpronik@crao.crimea.ua}

\altaffiltext{8}{Isaak Newton Institute of Chile,
	         Crimean Branch, Ukraine}

\altaffiltext{9}{Steward Observatory, 
		University of Arizona, 
		933 North Cherry Avenue, 
         	Tucson, AZ 85721; 
		mvestergaard@as.arizona.edu}

\altaffiltext{10}{Research Center for the Early Universe, 
		School of Science,
        	University of Tokyo, 
		7-3-1 Hongo, Bunkyo-ku, 
		Tokyo 113-0033, Japan}

\begin{abstract}

We describe results from a new ground-based monitoring campaign on
NGC~5548, the best studied reverberation-mapped AGN.  We find that it
was in the lowest luminosity state yet recorded during a monitoring
program, namely \mbox{$L_{\rm 5100} = 4.7 \times
10^{42}$~ergs~s$^{-1}$.}  We determine a rest-frame time lag between
flux variations in the continuum and the H$\beta$ line of
$6.3^{+2.6}_{-2.3}$ days.  Combining our measurements with those of
previous campaigns, we determine a weighted black hole mass of $M_{\rm
BH} = 6.54^{+0.26}_{-0.25} \times 10^7 M_{\odot}$ based on all broad
emission lines with suitable variability data.  We confirm the
previously-discovered virial relationship between the time lag of
emission lines relative to the continuum and the width of the emission
lines in NGC~5548, which is the expected signature of a
gravity-dominated broad-line region.  Using this lowest luminosity
state, we extend the range of the relationship between the luminosity
and the time lag in NGC~5548 and measure a slope that is consistent with
$\alpha = 0.5$, the naive expectation for the broad line region for an
assumed form of $r \propto L^{\alpha}$.  This value is also consistent
with the slope recently determined by \citeauthor{bentz06a} for the
population of reverberation-mapped AGNs as a whole.

\end{abstract}


\section{INTRODUCTION}

Reverberation mapping \citep{blandford82,peterson93} is an extremely
useful technique that has been exploited to measure the size of the
broad-line region (BLR) for many nearby active galactic nuclei (AGNs).
The time delay, or lag, between variations in the continuum flux and
broad line flux (usually H$\beta$) provides a measure of the size of the
region from which the broad lines emanate.  Coupling the measured lag
time with the velocity width of the emission line provides an estimate
of the mass of the central black hole.  To date, 36 nearby Seyfert
galaxies have BLR radius and black hole mass determinations from
reverberation mapping \citep{peterson04,peterson05}.

The Seyfert galaxy NGC~5548 was the focus of an intense, 13-year
campaign by the International AGN Watch
consortium\footnote{http://www.astronomy.ohio-state.edu/$\sim$agnwatch/}
\citep{peterson02} to study the variations in the optical continuum and
H$\beta$ line flux.  As a result, its emission line variability
properties are well characterized.

During the spring of 2005, we undertook a new ground-based monitoring
program (\citealt{bentz06b,denney06}) with the principal aim of
replacing H$\beta$ lag measurements and black hole mass estimates for
several AGNs with previous, unsatisfactory data.  We also monitored
NGC~5548 and found that it was in the lowest luminosity state yet
recorded during a monitoring campaign.  In this paper, we present the
low-luminosity monitoring data and combine it with previous monitoring
results of NGC~5548 to examine several questions related to
reverberation-mapping and AGN variability.  We explore the behavior of
the BLR with respect to varying luminosity states in NGC~5548 as well as
examine the intrinsic scatter in reverberation-based black hole mass
measurements for a single object.

\section{OBSERVATIONS AND DATA REDUCTION}

\subsection{Spectroscopy}

For most of the clear nights between 2005 March 1 and 2005 April 10, we
obtained spectra of the nucleus of NGC~5548 --- an S0/a Seyfert galaxy
at $z = 0.01717$ --- with the Boller and Chivens CCD Spectrograph (CCDS)
on the McGraw-Hill 1.3-m telescope at MDM Observatory. The observational
details and data reduction are as described by \citet{bentz06b}, but for
completeness, we include a few of the details here.  The observations of
NGC~5548 were obtained at a position angle of 90\degr\ through a
5\arcsec\ slit, resulting in a spectral resolution of 7.6~\AA\ over the
wavelength range $4400-5650$~\AA.  The typical seeing was 2\arcsec.

Additional spectra were obtained at the Crimean Astrophysical
Observatory (CrAO) 2.6-m Shajn Telescope with the Nasmith Spectrograph
and Astro-550 $580 \times 520$ pixel CCD \citep{berezin91}.  The spectra
were obtained through a 3\arcsec\ slit at a position angle of 90\degr.
Spectral reduction was carried out in the usual way with an extraction
width of 16 pixels, corresponding to 11\arcsec\ on the sky.

The spectra were then internally flux calibrated within each data set.
We scaled each individual spectrum to an [\ion{O}{3}] $\lambda 5007$
flux of $5.58 \times 10^{-13}$~ergs~s$^{-1}$~cm$^{-2}$ --- the flux of
the [\ion{O}{3}] $\lambda 5007$ line determined by photometric spectra
from the first year of the 13-year ground-based monitoring program for
NGC~5548 \citep{peterson91} --- using the spectral scaling algorithm of
\citet{vangroningen92}.  This process minimizes the residuals of the
[\ion{O}{3}] $\lambda 5007$ line in a difference spectrum by comparing
individual spectra to a reference spectrum (in this case, the mean
spectrum of all the data in the set) and scaling appropriately.

We then created light curves from measurements of the spectra. The mean
continuum flux was measured between observed-frame wavelengths of
5170--5200~\AA.  The flux of the H$\beta$ emission line was measured by
first setting a linear continuum level between the observed-frame
windows of 4800--4830~\AA\ and 5170--5200~\AA, and then integrating the
flux above the best-fit continuum between 4850--5000~\AA\ in the
observed frame.

\subsection{Photometry}

Photometry in the $V$-band was obtained at the 2.0-m Multicolor Active
Galactic NUclei Monitoring (MAGNUM) telescope at the Haleakala
Observatories in Hawaii \citep{kobayashi98b,yoshii02,yoshii03} with the
multicolor imaging photometer (MIP) \citep{kobayashi98a}.  Following the
procedure explained by \citet{suganuma04}, the nuclear flux of NGC~5548
was measured relative to a nearby reference star located at $(\Delta
\alpha, \Delta \delta) = (-0\farcm1, -2\farcm6)$. After processing the
frames in the usual way with IRAF, aperture photometry was performed
within an aperture radius of 4\farcs15, with sky subtraction between
radii of 5\farcs5--6\farcs9.  The reference star was then calibrated
using photometric standard stars from \citet{landolt92} and
\citet{hunt98}.  Variability of the reference star was previously shown
to be negligible by \citet{suganuma04}.  Due to technical issues with
the instrument, the photometric accuracy was very low at the beginning
of this monitoring campaign, but improved considerably towards the end
when the instrument was fixed (note the magnitude of the uncertainties
listed in Table~1 and the size of the error bars in Fig.~1).

\subsection{Inter-calibration of Light Curves}

As a result of the varying slit geometries, extraction apertures, and
seeing conditions in each of the spectroscopic and imaging data sets,
they must be inter-calibrated with each other to account for slit losses
and different amounts of host-galaxy starlight flux.  We follow the
process described by \citet{bentz06b} to scale the CrAO and MAGNUM data
sets to the MDM data set by using a least-squares analysis to identify
the flux offsets relative to the MDM data set.  After correcting for
these offsets, the three data sets were merged into a continuum light
curve and H$\beta$ light curve, which are tabulated in Table~1.  For the
analysis that follows, we combined observations within a 0.5 day window,
which yields the final light curves that are depicted in Figure~1.  The
points with the large error bars are those from the MAGNUM telescope
while the instrument was experiencing difficulties.

Table~2 describes the statistical properties of the final continuum and
emission-line light curves.  Column (1) lists the spectral feature of
the time series and column (2) gives the number of measurements.  The
mean and median sampling interval between data points are given in
columns (3) and (4), respectively.  The mean flux and standard deviation
of the time series are given in column (5).  The continuum flux
measurement does not take into account the host-galaxy starlight
contribution.  Using a high-resolution {\it Hubble Space Telescope}
image of NGC~5548 and the procedures described by \citet{bentz06a}, we
measure the host-galaxy contribution through the larger slit geometry in
this study ($5\farcs0~\times~12\farcs75$) to be $F_{\rm gal} ({\rm 5100
\AA}) = (5.29 \pm 0.49) \times
10^{-15}$~ergs~s$^{-1}$~cm$^{-2}$~\AA$^{-1}$.  The mean fractional
error, which is based on closely spaced observations in time, is given
in Column (6).  Column (7) is the excess variance, which is computed as

\begin{equation}
F_{\rm var} = \frac{\sqrt{\sigma^2 - \delta^2}}{\langle f \rangle}
\end{equation}

\noindent where $\sigma^2$ is the variance of the fluxes, $\delta^2$ is 
their mean-square uncertainty, and $\langle f \rangle$ is the mean of
the observed fluxes.  Lastly, column (8) is the ratio of the maximum to
the minimum flux in each time series.  Comparison of the values of
$F_{\rm var}$ for the continuum and H$\beta$ fluxes with those
determined for the previous 13 years of monitoring data show that the
amplitude of variation in the continuum is about half that of the
previous lowest continuum-variability year (Year 7 [1995] of
\citealt{peterson02}) and the amplitude of variability in the H$\beta$
line is the third lowest measured (see also Year 5 [1993] and Year 11
[1999] of \citeauthor{peterson02}).

Correcting the mean optical flux for the host-galaxy contribution above,
we find that NGC~5548 was in the lowest luminosity state yet recorded
during a monitoring campaign, only $F_{\rm AGN}(\rm{5100 \AA}) = (1.3
\pm 0.5) \times 10^{-15}$~ergs~s$^{-1}$~cm$^{-2}$~\AA$^{-1}$.  This is
a full 40\% fainter than the previous record holder, Year 4 (1992) of
the AGN Watch campaign (see \citealt{peterson02} for the original flux
measurement and \citealt{bentz06a} for the host galaxy correction).

\section{DATA ANALYSIS}

\subsection{Time Series}

As a result of the low luminosity of the AGN and the very low level of
variability throughout the campaign, it is quite evident that
determining a time lag can be problematic.  However, we proceed with the
usual tools developed for reverberation-mapping to see whether a
statistically significant detection can still be culled from the data.
To measure the time lag between the continuum and the broad part of the
H$\beta$ emission line, we employ the interpolation cross-correlation
function (ICCF) method of \citet{gaskell86} and \citet{gaskell87} and
the discrete correlation function (DCF) method of \citet{edelson88},
with the modifications discussed by \citet{white94} in both cases.  To
quantify the uncertainties in the time delay measurement, we follow the
method described by \citet{bentz06b}, originally outlined by
\citet{peterson98b} with the modifications of \citet{peterson02}.
Figure~1 shows the continuum and emission-line light curves and
cross-correlation results.

For the light curves of NGC~5548 shown in Figure~1, we find the
cross-correlation centroid occurs at $\tau_{\rm cent} =
6.4^{+2.6}_{-2.3}$~days and the peak of the cross-correlation function
(for physical lag times $\tau > 0$) occurs at $\tau_{\rm peak} =
6.5^{+2.5}_{-2.5}$~days in the observed frame.  Following the
consideration of the relative merits of $\tau_{\rm cent}$ versus
$\tau_{\rm peak}$ by \citet{peterson04}, we will use the value of
$\tau_{\rm cent}$ corrected for time dilation effects by a factor of
$(1+z)$, $\tau_{\rm cent} = 6.3^{+2.6}_{-2.3}$~days, in the discussion
that follows.

A measure of the quality of the data can be estimated by analysis of the
auto-correlation function of the continuum in the top-right panel of
Figure~1.  The ICCF and DCF methods give very different values at $\tau
= 0$, the peak of the ICCF is $1.0$, compared with the peak of the DCF,
$\sim 0.5$.  This discrepancy arises from the fact that we have
discarded measurements of the DCF where $\tau$ is exactly equal to 0
(i.e., simultaneous pairs measured from the same spectrum) because they
are subject to correlated errors, especially when the amplitude of
variability is so low.  The DCF with zero-lag pairs excluded therefore
gives a more realistic measure of the correlation of nearby data points
with each other because, unlike the ICCF, it does not include
correlation of the noise in the data with itself.  With such noisy data
and low-level signals in the light curves, it is rather remarkable that
we are able to obtain a statistically significant time lag between the
continuum and H$\beta$ light curves.  

To further quantify the significance of this detection, we have carried
out the following Monte Carlo simulations.  An optical continuum
light curve was generated using the characteristics of the power density
spectrum measured for NGC~5548, $P(f) \propto f^{-2.56}$
\citep{collier01}.  The model continuum light curve was then convolved
with a transfer function to produce a model emission-line light curve.
Two simple models for the transfer function were employed for
illustrative purposes:

\begin{itemize}

\item{A thin, spherical shell of radius $R = 6.3$~light days, which has 
a constant response from $\tau = 0$ to $\tau = 2R/c$ and is zero
everywhere else.  Such a transfer function results in the most smoothing
of the continuum light curve shape as it is transferred to the line
light curve.}

\item{A thin disk of radius $R = 6.3$~light days and inclination 0\degr, 
which is a delta function at $\tau = R/c$.  This transfer function
results in the sharpest transfer of structural features from the
continuum light curve to the emission-line light curve.}

\end{itemize}

\noindent  For ideal data, each of these cases would yield a lag 
measurement of $\tau = R/c$.

After creating the model light curves, they were sampled in the same
pattern as the observations presented in this paper.  The sampled points
were rescaled according to the excess variances and mean fractional
errors determined for the real data set and presented in Table~2.  To
simulate noise, each data point was randomly altered by a Gaussian
deviate within the flux uncertainties appropriate for the observing
campaign.  The simulated set of observations was then cross-correlated
in the manner described above, and the peak and centroid of the cross
correlation was recorded.  This entire process, beginning with the model
continuum generation, was repeated 1000 times for each of the transfer
functions described above.  The resulting cross-correlation peak and
centroid distributions for the two transfer functions are displayed in
Figure~2.  Casual inspection of the distributions in Figure 2 shows that
most of the simulation results pile up around the expected lag
measurement of 6.3 days, which lends much confidence to the measurements
made above for our observing campaign.

About 15\% of the simulations result in unphysical lag times, that is,
$\tau < 0$, indicating that the broad line flux responds before the
continuum flux varies.  Considering only those simulations that have
physically plausible lag times ($\tau \geq 0$), we find that for the
thin shell transfer function, 53\% of the peak measurements and 56\% of
the disk measurements fall within the range of $1\sigma$ uncertainties
quoted above for the lag measured in our observing campaign.  The
percentages increase to 69\% and 67\%, respectively, for the thin disk
transfer function.  The asymmetric shapes of the peak and centroid
distributions result in slightly lower median lag times than expected,
however the medians are well within $1\sigma$ of the input lag time of
6.3~days: $\tau_{\rm peak} = 5.8^{+3.0}_{-3.2}$ and $\tau_{\rm cent} =
5.5^{+2.8}_{-3.0}$, based on the simulations with a thin shell transfer
function; $\tau_{\rm peak} = 6.0^{+1.8}_{-2.5}$ and $\tau_{\rm cent} =
5.8^{+1.7}_{-2.8}$, based on the simulations with a thin disk transfer
function.  The fact that the thin disk model seems to more accurately
reproduce the observed data is not surprising: \citet{horne91} find
strong evidence ruling out a spherically symmetric geometry for the
H$\beta$ BLR of NGC~5548.  The most likely geometry is probably
somewhere in between these two simple models, however, the models are
useful in that they approximate the two opposite extremes of possible
BLR behavior in AGNs in the sense of minimal (face-on disk) versus
maximal (spherical shell) smoothing of the line light curve shape.  In
any case, the simulations clearly confirm the detected restframe lag of
6.3~days.

\subsection{Line Width Measurement}

The mean and root-mean-square (RMS; i.e. variable) spectra of NGC~5548
were calculated using the full set of MDM data and are shown in
Figure~3.  To measure the width of the broad emission line, we first
subtract the [\ion{O}{3}] $\lambda 4959$ narrow line using the
[\ion{O}{3}] $\lambda 5007$ emission line as a template and the standard
scaling factor of 0.340 for the $\lambda 4959$ line relative to the
$\lambda 5007$ line \citep{storey00}.  We also subtract the narrow
component of the H$\beta$ line using the same template and the scaling
value of 0.110 for the ratio of the lines determined by
\citet{peterson02}.  We then interpolate the continuum underneath the
broad H$\beta$ emission line by setting continuum windows on either side
of the line.  The width of the line is typically characterized by its
full width at half maximum (FWHM) and by the second moment of the line
profile --- the line dispersion ($\sigma_{\rm line}$) --- as described
by \citet{peterson04}, and is corrected for the spectral resolution.  In
this case, since the red side of the mean profile can be contaminated by
\ion{Fe}{2} multiplet 42 emission and artifacts of the [\ion{O}{3}]
removal, we also include in our line measurements the line dispersion
based on the blue side of the line, assuming the profile is
approximately symmetric around the expected central wavelength of the
line, which we denote $\sigma_{\rm blue,sym}$.  The uncertainties in the
line width measurements are determined using the method described by
\citet{bentz06b}, however, the formal uncertainties for the $\sigma_{\rm
line}$ and $\sigma_{\rm blue,sym}$ measurements in the mean spectrum of
0.6\% are probably misleading since the measurement errors are mostly
systematic in this low state of variability.  We therefore adopt an
error of 20\%, which is more typical of the errors in mean profile
widths.

Table~3 lists the line widths measured for NGC~5548 in the mean
spectrum, where the signal is much stronger, as well as the RMS
spectrum.  For this particular data set, the FWHM of the variable (RMS)
broad line is not well characterized, so we omit this measurement.
Fortunately, all measurements of the line width tabulated in Table~3 are
consistent with each other even though the RMS spectrum in particular
has an extremely low-level signal.  While the line width measurement
that is typically preferred for reverberation-mapping data sets is
$\sigma_{\rm line}$(RMS) \citep{collin06}, due to the low-level signal
in the RMS spectrum, we will choose to use a measurement from the mean
spectrum in this case.  The blending problems between the red side of
the H$\beta$ line and the [\ion{O}{3}] $\lambda 4959$ line and other
features cause us to adopt $\sigma_{\rm blue,sym}$(mean) as the most
credible line width for this data set in the following discussion.
While we have noted that all the line width measurements for this data
set are consistent with each other, so the choice of line width does not
have a critical impact on the following discussion, there are issues
that effect certain measurements and not others, so a combination of the
line width measurements is not appropriate here.

\subsection{Black Hole Mass}

The mass of the central black hole is determined by

\begin{equation}
M_{\rm BH} = \frac{f c \tau \Delta V^2}{G},
\end{equation}

\noindent where $\tau$ is the time delay of the emission line, $\Delta V$ 
is the width of the emission line, $c$ is the speed of light, and $G$ is
the gravitational constant.  The factor $f$ takes into account the
unknown inclination, geometry, and kinematics of the BLR.
\citet{onken04} have found that an average value $\langle f \rangle =
5.5 \pm 1.7$ if the relationship between black hole mass and stellar
velocity dispersion ($M_{\rm BH} - \sigma_*$) for AGNs is normalized to
that of quiescent galaxies.  With this normalization, we find $M_{\rm
BH} = 7.0^{+4.0}_{-3.7} \times 10^7 M_{\odot}$.  Combining this new
measurement with all previous reverberation-based mass measurements of
the black hole in NGC~5548, we find a weighted mean of $M_{\rm BH} =
6.83^{+0.30}_{-0.28} \times 10^7 M_{\odot}$ based on H$\beta$
measurements only, and $M_{\rm BH} = 6.54^{+0.26}_{-0.25} \times 10^7
M_{\odot}$ when including all measurements of broad lines.  The
reverberation-based mass is in good agreement with the mass expected
from the $M_{\rm BH} - \sigma_*$ relationship of \citet{tremaine02},
which predicts $M_{\rm BH} = 9.4^{+7.6}_{-4.2} \times 10^7 M_{\odot}$
based on the stellar velocity dispersion measurement for NGC~5548 of
$\sigma_* = 183 \pm 27$~km~s$^{-1}$ \citep{ferrarese01}.  Considering
the low-level signals of both the time lag and the line width in this
low-luminosity data set, it is rather remarkable that our results are
consistent with the expectations based on previous monitoring programs
as well as other methods of estimating the black hole mass for NGC~5548.

\section{DISCUSSION}

With the addition of this lowest-luminosity data set to the compilation
of NGC~5548 monitoring data, we examine several relationships between
luminosity and other parameters where this experiment extends the range
that may be explored.

\subsection{Relationship Between Line Width and Time Lag}

Figure~4 shows the width of broad emission lines ($\sigma_{\rm line}$)
measured in NGC~5548 versus their time lag relative to the continuum at
5100~\AA\ ($\tau_{cent}$).  If the BLR kinematics are dominated by
gravity, we expect that there will exist a virial relationship between
the time lag and the width of an emission line.  The top panel shows the
relationship for only those measurements of the H$\beta$ line in
NGC~5548, and the bottom panel includes all broad lines available.  In
either case, the slope is basically the same, about $-0.5 \pm 0.1$,
consistent with the virial relationship previously reported for NGC~5548
\citep{peterson99,peterson04}.

\subsection{Relationship Between Time Lag and Luminosity}

Photoionization models of the BLR are typically parameterized by the
shape of the ionizing continuum incident on the BLR gas, and an
ionization parameter $U$, defined as

\begin{equation}
U = \frac{Q(H)}{4 \pi r^2 n_{\rm H} c},
\end{equation}

\noindent where $Q(H)$ is the number of ionizing photons emitted from the 
continuum source per second, $r$ is the size of the BLR, and $n_{\rm H}$
is the BLR particle density.  We would naively expect that as the
ionizing continuum luminosity varies, each emission line should have the
greatest response at some particular value of $U n_{\rm H}$.  This would
then lead to the expectation that, for any given line, $r \propto L_{\rm
ion}^{1/2}$.

In Figure~5 we plot the relationship between the H$\beta$ time lag
relative to the continuum and the luminosity of the AGN at 5100 \AA.  If
we include all 13 years of data from the AGN Watch campaign and the new
low-luminosity data supplied in this work, we find a power law slope of
$0.73 \pm 0.14$.  However, the data from Year 12 (2000) of the AGN Watch
program is the most poorly sampled of all the monitoring campaigns for
NGC~5548.  The cross-correlations of the H$\beta$ emission line with the
continuum in the Year 12 data yields very ambiguous results, as can
easily be seen in Figure~2 of \citet{peterson02}.  If we remove the Year
12 data point based on its potential unreliability, we find a power law
slope of $0.66 \pm 0.13$, within $\sim 1 \sigma$ of the naive
expectation of 0.5.  Using this relationship and the mean luminosity of
NGC~5548, we estimate that the lag time for the Year 12 campaign should
have been measured at $\sim 11$~days rather than $\tau_{\rm cent} =
6.6^{+5.8}_{-3.8}$~days as determined by \citet{peterson04}.  With the
large error bars due to the ambiguous results of the cross-correlations
for Year 12, the expected value of 11~days is within $1 \sigma$ of the
measured value of 6.6~days.  Based on these arguments, we are more
inclined to accept the slope of $0.66 \pm 0.13$.

One would expect that a measurement of the AGN luminosity in the
ultraviolet (UV) would give a much better estimate of $L_{\rm ion}$ than
in the optical part of the spectrum.  In Figure~6, we examine the
relationship between 83 pairs of optical flux measurements at 5100~\AA,
$F_{\rm opt}$~(5100~\AA), and UV flux measurements at 1350~\AA, $F_{\rm
UV}$~(1350~\AA), for NGC~5548.  The UV and optical flux measurements in
each pair were made within one day of each other, and no statistically
significant time delay exists between the optical and UV data (see
\citealt{peterson02} for a discussion of the data).  The optical fluxes
have been corrected using the host-galaxy starlight contribution
measured by \citet{bentz06a}.  The best power law fit to the data is
$F_{\rm opt} \propto F_{\rm UV}^{0.84 \pm 0.05}$.  If we combine this
with the above $r - L_{\rm opt}$ fit excluding the data from Year 12, we
find $r \propto L_{\rm UV}^{0.55 \pm 0.14}$.  Interestingly, this is
consistent not only with the naive expectation of how an object should
behave when it varies over time, but also with the relationship recently
determined by \citet{bentz06a} for the population of
reverberation-mapped AGNs as a whole.  On the other hand,
\citet{cackett06} find a much shallower slope when modeling the 13 years
of AGN Watch data for NGC~5548 as one continuous light curve with a
luminosity-dependent delay map, more in agreement with the predictions
of detailed photoionization models by \citet{korista04}.  However,
\citeauthor{cackett06} used a smaller correction for the host galaxy
starlight contribution to the flux, which would serve to artificially
flatten the relationship they find.

\subsection{Black Hole Mass Estimates}

The most obvious relationship left to examine is whether the black hole
mass determinations are changing with the luminosity of the AGN.  We
expect that this would not be the case, as the black hole mass should
not be changing in a measurable way and it certainly should not depend
on the luminosity.  Figure~7 shows the distribution of virial products
(the black hole mass without the average scaling factor of 5.5, i.e.,
$M_{\rm BH}/f$) as a function of luminosity.  This is similar to the
recent analysis by \citet{collin06} in that the open circles are those
based on the measurement of $\sigma_{\rm blue,sym}$ from the mean
spectrum, and the filled circles are based on the measurement of
$\sigma_{\rm line}$ from the RMS spectrum.  The luminosity measurements
have been corrected for the contribution from host-galaxy starlight
using the corrections of \citet{bentz06a}.  We have added the lowest
luminosity data point described in this work to the set previously
examined by \citeauthor{collin06}, but we use $\sigma_{\rm blue,sym}$
from both the mean and RMS spectra because of the blending issues
discussed earlier in $\S 3.2$.

Inspection of Figure~7 shows that the virial product for NGC~5548 is
consistent with being independent of luminosity state.  This is as
expected, since the black hole mass should not be varying with
luminosity.  The largest outlier is again from Year 12 (2000) of the AGN
Watch program \citep{peterson02}.

Formal likelihood analysis of the black hole mass measurements gives a
width of 0.15~dex, which is comparable to the standard deviation of the
measurements themselves.  From this, we may conclude that the intrinsic
scatter dominates the errors for repeated reverberation-based mass
measurements of a single object.

While the reverberation-mapping experiment appears to be quite
repeatable with a fairly high precision, we do not have an independent
measurement of the systematic errors present in reverberation-mapping
techniques.  In order to estimate the systematic errors, we must compare
results with a complementary method of measuring $M_{\rm BH}$.
Comparison of reverberation-mapping results to those derived from the
$M_{\rm BH} - \sigma_*$ relationship shows that there is a factor of
$\sim 3$ scatter in $M_{\rm BH}$ measurements from reverberation-mapping
\citep{onken04}.  It is therefore important that inclination and other
effects be considered in order to reduce the systematic errors in
reverberation-mapping results.

\section{CONCLUSIONS}
We present the lowest-luminosity ($L_{5100} = 4.7 \times 10^{42}$ ergs
s$^{-1}$) monitoring data for NGC~5548. We measure a rest-frame time lag
of $6.3^{+2.6}_{-2.3}$ days for changes in the H$\beta$ line flux
relative to changes in the continuum flux. Combining this new data point
with previously collected data, we find the weighted mean of the black
hole mass to be $M_{\rm BH} = 6.54^{+0.26}_{-0.25} \times 10^7
M_{\odot}$.

With the results of this low-luminosity monitoring campaign, we are able
to extend the range over which we may examine several relationships that
depend on luminosity.  We confirm the existence of a virial relationship
between the time lag and the line width of the broad emission lines in
NGC~5548.  We find that the relationship between luminosity and time lag
in NGC~5548 is consistent with a virial expectation, and is also
consistent with the relationship determined for the entire population of
AGNs with mass measurements from reverberation mapping.

\acknowledgements
We thank the referee, Bill Welsh, for suggestions that improved the
presentation of this paper.  We also thank Andy Gould for helpful
comments and suggestions.  We are grateful for support of this work by
the National Science Foundation through grants AST-0205964 and
AST-0604066 to The Ohio State University and the Civilian Research and
Development Foundation through grant UP1-2549-CR-03.  MCB is supported
by a Graduate Fellowship of the National Science Foundation.  KDD is
supported by a GK-12 Fellowship of the National Science Foundation.  EMC
gratefully acknowledges support from PPARC.  MV acknowledges financial
support from NSF grant AST-0307384 to the University of Arizona.  This
research has made use of the NASA/IPAC Extragalactic Database (NED)
which is operated by the Jet Propulsion Laboratory, California Institute
of Technology, under contract with the National Aeronautics and Space
Administration.


\clearpage

\begin{deluxetable}{cccc}
\tablecolumns{4}
\tablewidth{0pt}
\tablecaption{Continuum and H$\beta$ Fluxes for NGC~5548}
\tablehead{
\colhead{JD\tablenotemark{a}} &
\colhead{F$_{\lambda}$ (5100 \AA)} &
\colhead{H$\beta$ $\lambda 4681$} &
\colhead{Observatory}\\
\colhead{(-2,450,000)} &
\colhead{($10^{-15}$ ergs s$^{-1}$ cm$^{-2}$ \AA$^{-1}$)} &
\colhead{($10^{-13}$ ergs s$^{-1}$ cm$^{-2}$)} &
\colhead{Code\tablenotemark{b}}}

\startdata

3431.996 & 6.15 $\pm$ 0.18 & 1.79   $\pm$ 0.05   & M \\
3432.922 & 5.27 $\pm$ 0.52 & \nodata             & H \\
3435.930 & 6.63 $\pm$ 0.39 & \nodata		 & H \\
3437.969 & 6.80 $\pm$ 0.20 & \nodata		 & H \\
3438.023 & 6.56 $\pm$ 0.19 & 2.04   $\pm$ 0.06   & M \\
3438.922 & 6.74 $\pm$ 0.20 & 2.21   $\pm$ 0.07   & M \\
3440.996 & 6.75 $\pm$ 0.20 & 2.19   $\pm$ 0.07   & M \\
3441.980 & 6.64 $\pm$ 0.19 & 2.27   $\pm$ 0.07   & M \\
3442.891 & 6.52 $\pm$ 0.30 & \nodata		 & H \\
3443.004 & 6.48 $\pm$ 0.19 & 2.26   $\pm$ 0.07   & M \\
3443.996 & 6.52 $\pm$ 0.19 & 2.23   $\pm$ 0.07   & M \\
3444.598 & 6.31 $\pm$ 0.30 & 2.30   $\pm$ 0.04   & C \\
3445.594 & 6.13 $\pm$ 0.29 & 2.35   $\pm$ 0.04   & C \\
3446.008 & 6.65 $\pm$ 0.19 & 2.29   $\pm$ 0.07   & M \\
3447.000 & 6.45 $\pm$ 0.19 & 2.38   $\pm$ 0.07   & M \\
3447.000 & 6.26 $\pm$ 0.17 & \nodata		 & H \\
3450.973 & 6.54 $\pm$ 0.19 & 2.26   $\pm$ 0.07   & M \\
3451.922 & 6.68 $\pm$ 0.19 & \nodata		 & H \\
3451.941 & 6.51 $\pm$ 0.19 & 2.41   $\pm$ 0.07   & M \\
3452.957 & 6.74 $\pm$ 0.20 & 2.22   $\pm$ 0.07   & M \\
3456.949 & 7.08 $\pm$ 0.21 & 2.07   $\pm$ 0.06   & M \\
3457.852 & 6.54 $\pm$ 0.67 & \nodata		 & H \\
3459.004 & 6.88 $\pm$ 0.20 & 2.48   $\pm$ 0.07   & M \\
3459.922 & 6.87 $\pm$ 0.20 & 2.39   $\pm$ 0.07   & M \\
3460.938 & 6.75 $\pm$ 0.20 & 2.51   $\pm$ 0.08   & M \\
3461.887 & 7.25 $\pm$ 0.21 & 2.47   $\pm$ 0.07   & M \\
3462.918 & 6.73 $\pm$ 0.20 & 2.46   $\pm$ 0.07   & M \\
3463.527 & 6.99 $\pm$ 0.34 & 2.48   $\pm$ 0.04   & C \\
3464.504 & 6.86 $\pm$ 0.33 & 2.45   $\pm$ 0.04   & C \\
3464.891 & 6.69 $\pm$ 0.19 & 2.43   $\pm$ 0.07   & M \\
3465.949 & 6.58 $\pm$ 0.19 & 2.51   $\pm$ 0.08   & M \\
3466.070 & 7.04 $\pm$ 0.09 & \nodata		 & H \\
3466.922 & 7.33 $\pm$ 0.21 & 2.52   $\pm$ 0.08   & M \\
3467.973 & 6.52 $\pm$ 0.19 & 2.50   $\pm$ 0.08   & M \\
3468.957 & 6.45 $\pm$ 0.19 & 2.55   $\pm$ 0.08   & M \\
3469.922 & 6.45 $\pm$ 0.19 & 2.42   $\pm$ 0.07   & M \\
3470.012 & 6.49 $\pm$ 0.09 & \nodata		 & H \\
3470.926 & 6.53 $\pm$ 0.19 & 2.48   $\pm$ 0.07   & M \\
3471.516 & 6.73 $\pm$ 0.32 & 2.31   $\pm$ 0.04   & C \\
3471.922 & 6.34 $\pm$ 0.18 & 2.46   $\pm$ 0.07   & M \\

\enddata

\tablenotetext{a}{The Julian Date listed is the midpoint of the observation}.

\tablenotetext{b}{Observatory Codes:
		  {\bf C} = CrAO 2.6-m Shajn Telescope + Nasmith 
				Spectrograph;
		  {\bf H} = Haleakala Observatories 2.0-m MAGNUM Telescope 
				+ MIP;
		  {\bf M} = MDM Observatory 1.3-m McGraw-Hill Telescope + 
				CCDS.}
\end{deluxetable}

\begin{deluxetable}{lccccccc}
\tablecolumns{8}
\tablewidth{0pt}
\tablecaption{Light Curve Statistics}
\tablehead{
\colhead{} &
\colhead{} &
\multicolumn{2}{c}{Sampling} &
\colhead{} &
\colhead{Mean} &
\colhead{} &
\colhead{} \\
\colhead{Time} &
\colhead{} &
\multicolumn{2}{c}{Interval (days)} &
\colhead{Mean} &
\colhead{Fractional} &
\colhead{} &
\colhead{} \\
\colhead{Series} &
\colhead{$N$} &
\colhead{$\langle T \rangle$} &
\colhead{$T_{\rm median}$} &
\colhead{Flux\tablenotemark{a}} &
\colhead{Error} &
\colhead{$F_{\rm var}$} &
\colhead{$R_{\rm max}$}\\
\colhead{(1)} &
\colhead{(2)} &
\colhead{(3)} &
\colhead{(4)} &
\colhead{(5)} &
\colhead{(6)} &
\colhead{(7)} &
\colhead{(8)}}
\startdata

5100 \AA & 31 & 1.3 & 1.0 & $6.63 \pm 0.36$ & 0.033 & 0.040 & $1.39 \pm 0.14$ \\ 
H$\beta$ & 28 & 1.5 & 1.0 & $2.34 \pm 0.17$ & 0.027 & 0.069 & $1.42 \pm 0.06$ \\

\enddata
\tablenotetext{a}{Fluxes are in the same units as those used in Table~1.}
\end{deluxetable}

\clearpage

\begin{deluxetable}{lcc}
\tablecolumns{3}
\tablewidth{0pt}
\tablecaption{H$\beta$ Line Width Measurements}
\tablehead{
\colhead{Line} &
\colhead{Mean} &
\colhead{RMS}\\
\colhead{Width} &
\colhead{Spectrum} &
\colhead{Spectrum}\\
\colhead{Measurement} &
\colhead{(km s$^{-1}$)} &
\colhead{(km s$^{-1}$)}}
\startdata

$\sigma_{\rm line}$ 	  & $2662 \pm 532$  		     & $2388 \pm 373$ \\
$\sigma_{\rm blue, sym}$  & $3210 \pm 642$\tablenotemark{a}  & $2939 \pm 768$ \\
FWHM			  & $6396 \pm 167$                   & \nodata \\

\enddata
\tablenotetext{a}{The preferred measurement of the width of H$\beta$ for the
		  new data presented in this paper.} 
\end{deluxetable}

\begin{figure}
\figurenum{1}
\epsscale{1}
\plotone{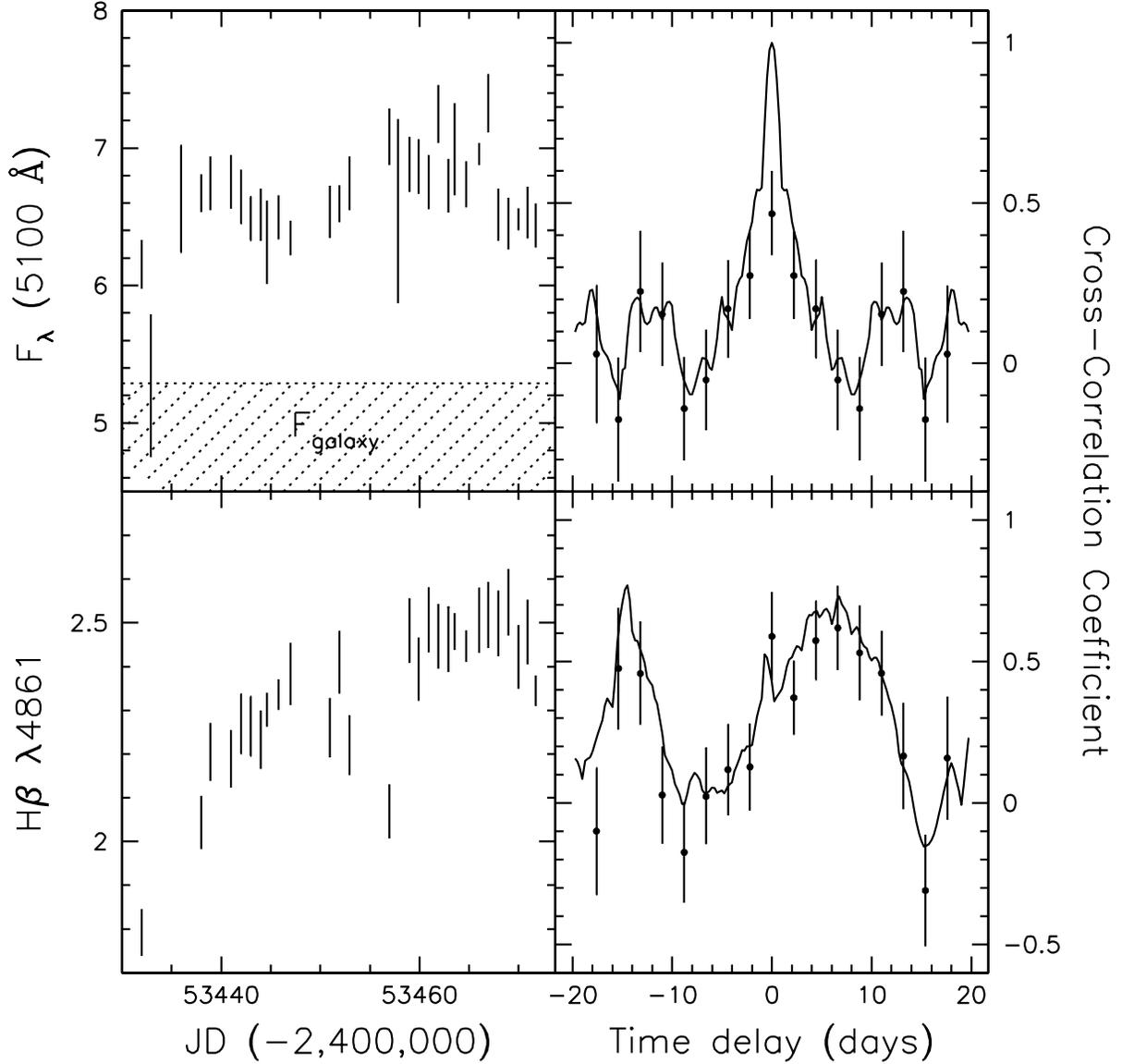}
\caption{The left panels show the light curves for the continuum region 
         at 5100 \AA\ and for the H$\beta$ $\lambda 4861$ line.
         Measurements taken within a 0.5 day bin have been averaged
         together. The flux is in units of $10^{-15}$ ergs s$^{-1}$
         cm$^{-2}$ \AA$^{-1}$ for the continuum and units of $10^{-13}$
         ergs s$^{-1}$ cm$^{-2}$ for H$\beta$.  The shaded area of the
         continuum light curve shows the contribution to the flux from
         the host galaxy.  The right panels show the result of
         cross-correlating each light curve with the continuum light
         curve; the top right panel is therefore the continuum
         auto-correlation function.  The solid line shows the ICCF
         method, while the points show the DCF method, as described in
         the text.}
\end{figure}

\clearpage

\begin{figure}
\figurenum{2}
\epsscale{1}
\plotone{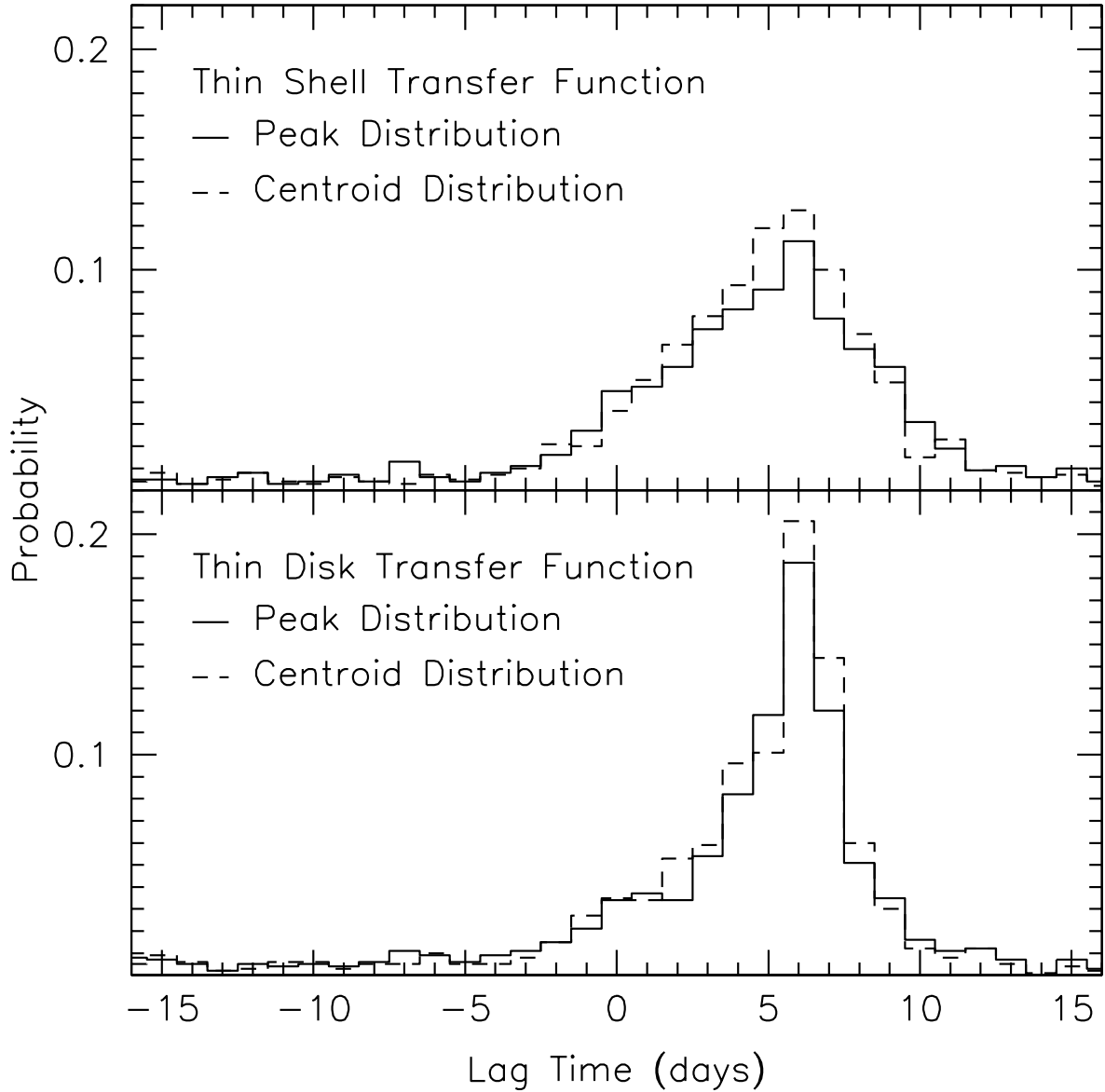}
\caption{Cross-correlation peak distributions (solid lines) and centroid 
         distributions (dashed lines) for the Monte Carlo simulations
         described in the text.  The top panel is for a thin shell
         transfer function, and the bottom panel is for a thin disk
         transfer function.}
\end{figure}

\clearpage

\begin{figure}
\figurenum{3}
\epsscale{1}
\plotone{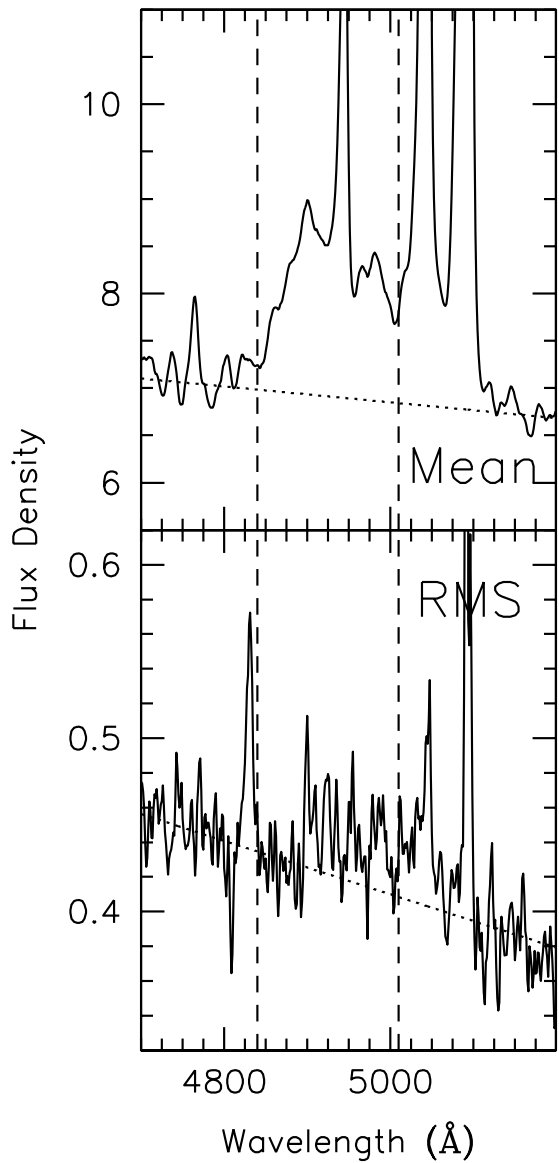}
\caption{Mean and RMS of the MDM spectra in the observed frame of NGC~5548.  
         The RMS spectrum still shows contributions from the
         [\ion{O}{3}] $\lambda 4959$ and $\lambda 5007$ lines due to
         imperfect subtraction of all the nightly line profiles.  The
         spike at 4825~\AA\ appears to be due to a bad column in the
         detector.  The dashed lines in both windows show the wavelength
         limits and the dotted lines show the continuum levels used in
         calculating the width of the broad H$\beta$ line.  The flux
         density for both spectra is in units of $10^{-15}$ ergs
         s$^{-1}$ cm$^{-2}$ \AA$^{-1}$.}
\end{figure}

\clearpage

\begin{figure}
\figurenum{4}
\epsscale{1}
\plotone{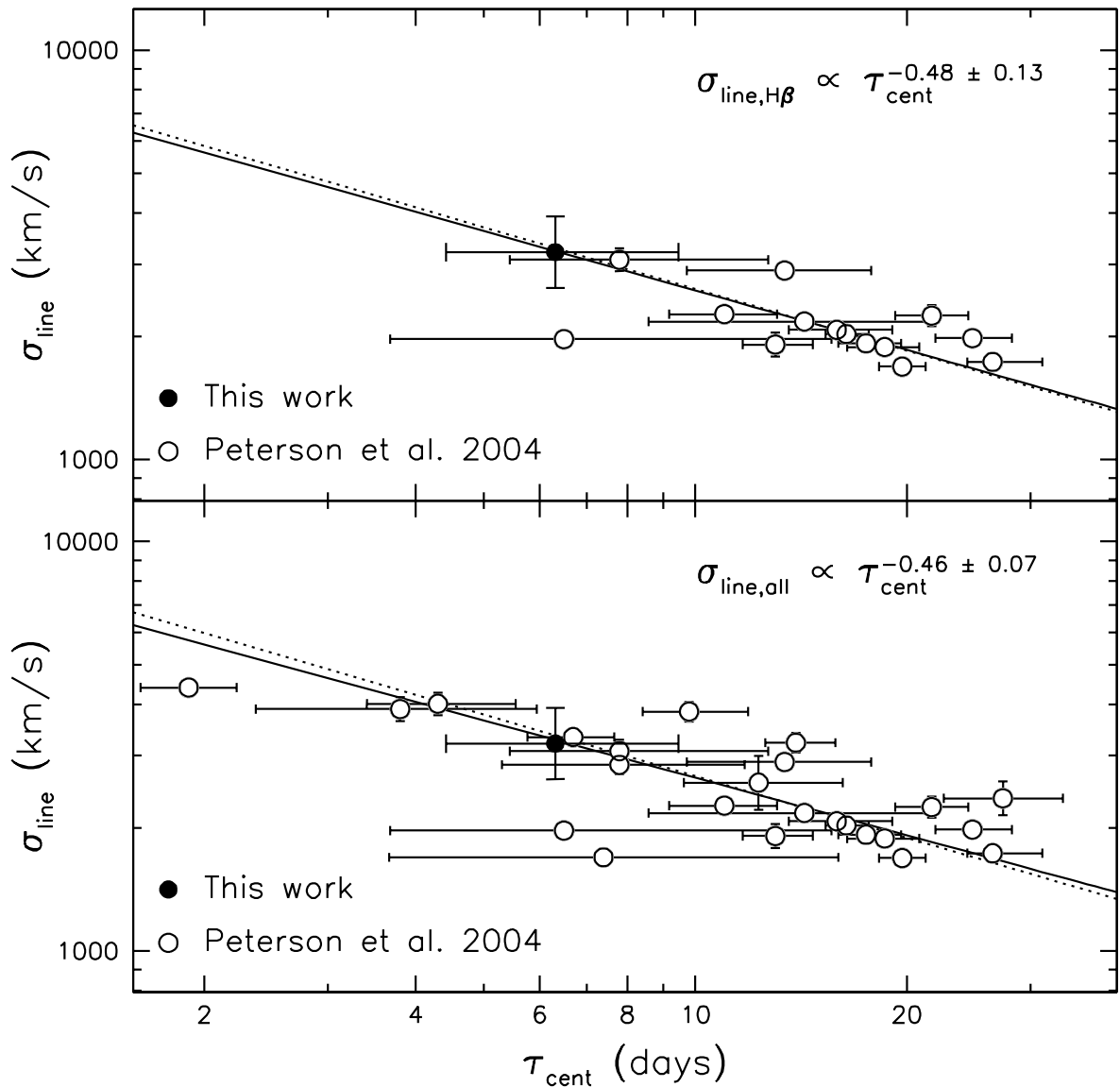}
\caption{Rest-frame emission line widths versus time lags for NGC~5548.  
         The top box shows the relationship for all measurements of the
         H$\beta$ line, while the bottom box shows the relationship for
         measurements of all lines.  The filled circles show the
         measurement of H$\beta$ from this work, while the open circles
         are from \citet{peterson04}.  The solid lines are the best fits
         to the data, and the dotted lines are fits with a forced slope
         of -0.5, i.e. a virial relationship.}
\end{figure}

\clearpage

\begin{figure}
\figurenum{5}
\epsscale{1}
\plotone{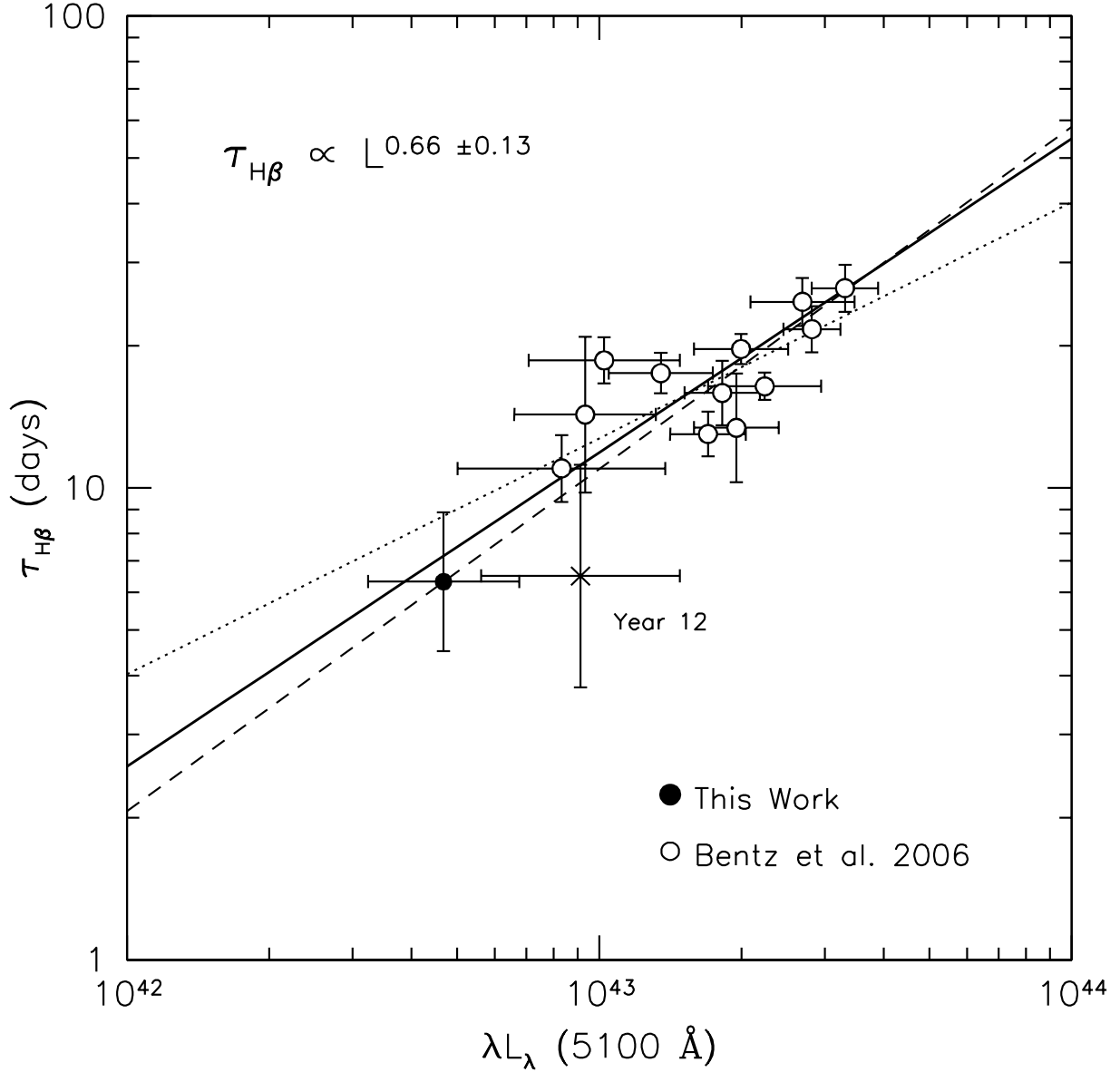}
\caption{Rest-frame H$\beta$ time lag versus the luminosity at 5100 \AA.  
         The open circles are from \citet{bentz06a} after the luminosity
         measurement has been corrected for the host galaxy starlight
         contribution, and the filled circle is from this work, also
         corrected for host galaxy starlight.  The dashed line has a
         slope of $0.73 \pm 0.14$ and shows the fit to the data with all
         the points included.  The solid line shows the fit excluding
         the data from Year 12, with a slope of $0.66 \pm 0.13$,
         which we take to be the best current fit to the data for
         NGC~5548.  The dotted line has a forced slope of 0.5, as one
         would expect from naive predictions of the behavior of the gas
         in the BLR to changes in the ionizing flux.}
\end{figure}

\clearpage

\begin{figure}
\figurenum{6}
\epsscale{1}
\plotone{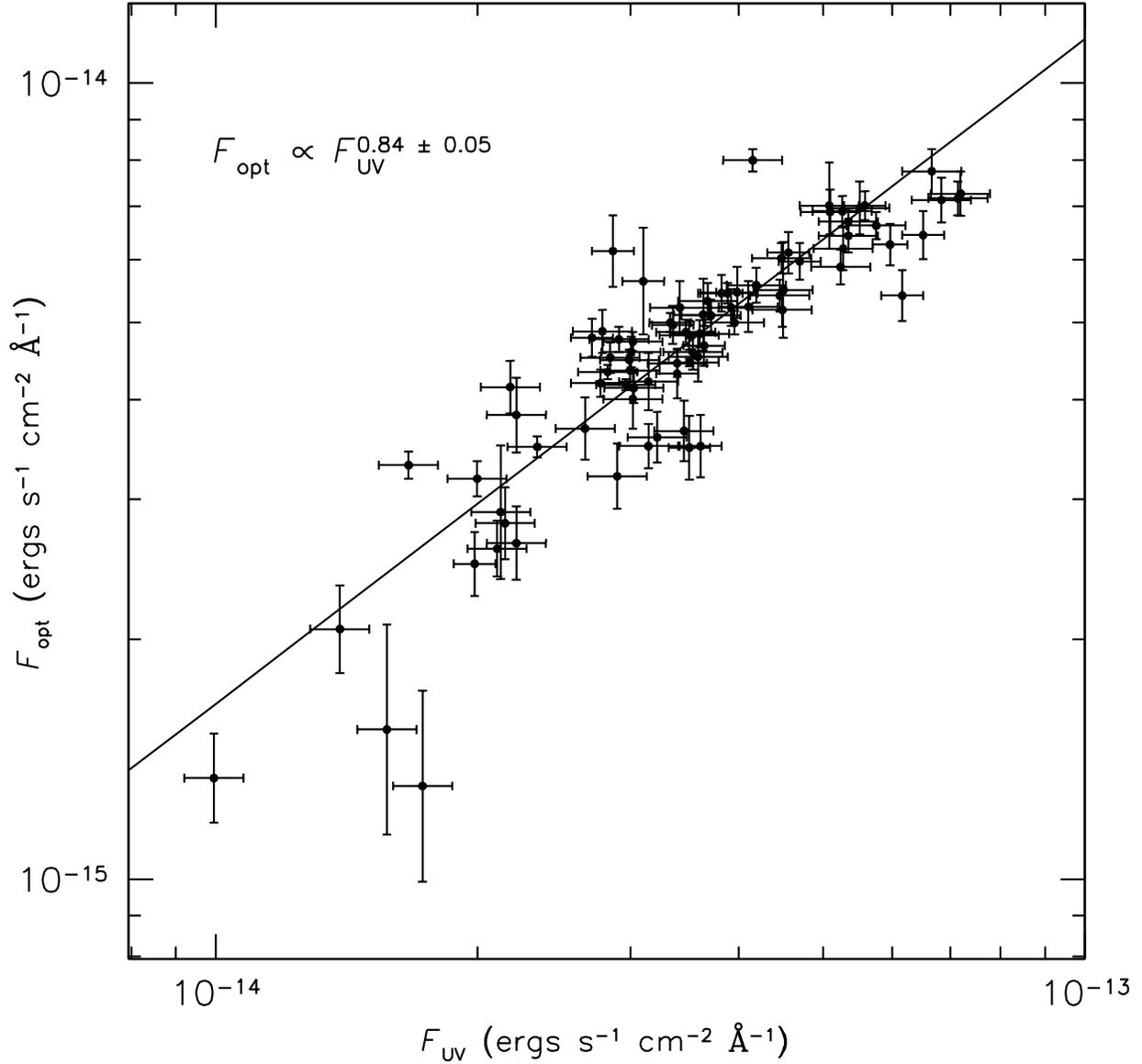}
\caption{Optical flux at 5100~\AA, after correction for host galaxy starlight, 
         versus the ultraviolet flux at 1350~\AA.  Each pair of optical
         and UV measurements was made within one day.  The solid line
         shows the best fit, which has a power law slope of $0.84 \pm
         0.05$.}
\end{figure}

\clearpage

\begin{figure}
\figurenum{7}
\epsscale{1}
\plotone{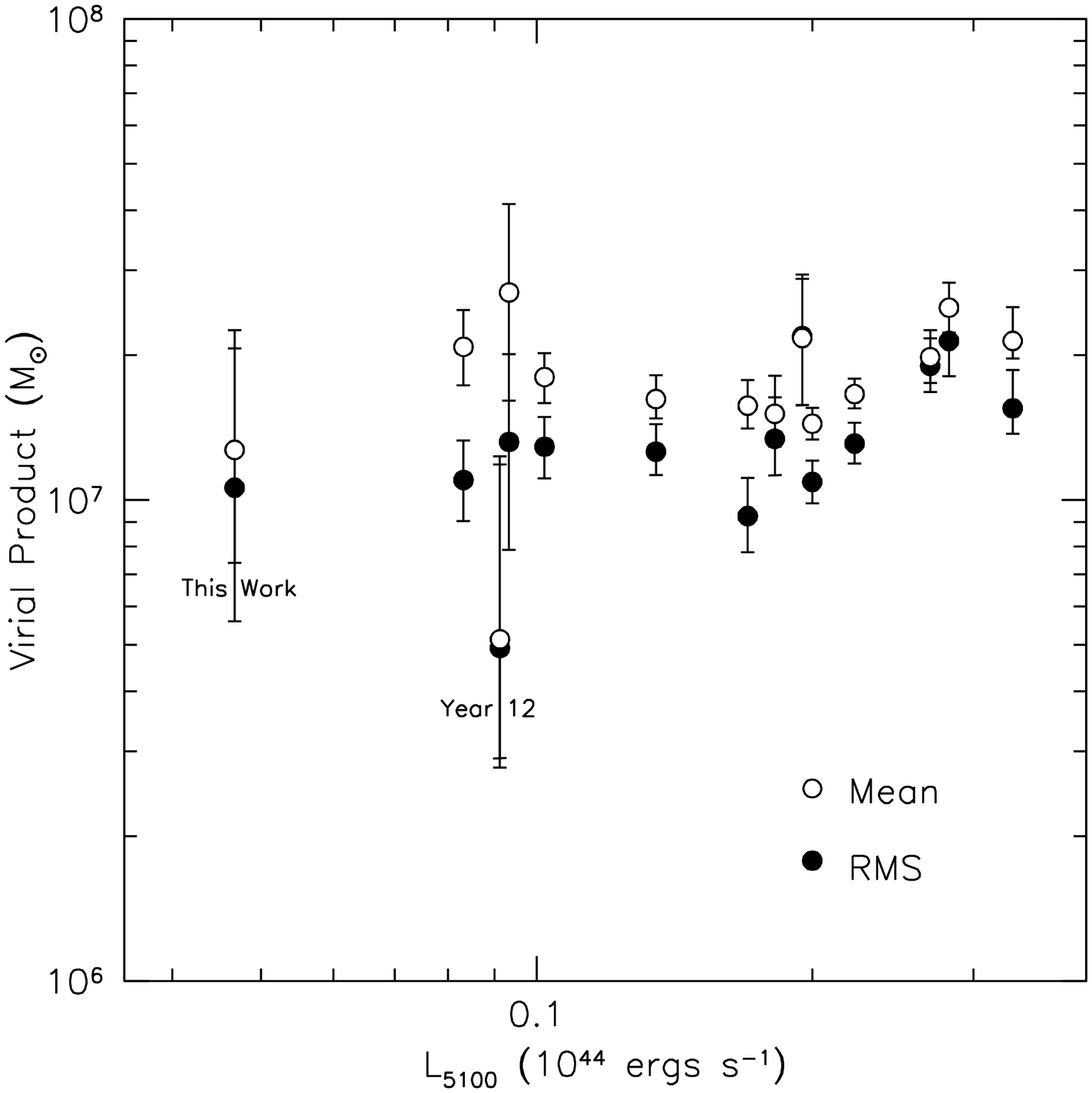}
\caption{The virial product determined with $\sigma_{\rm line}$ calculated 
	 from the mean spectrum (open circles) and the RMS spectrum
	 (filled circles) versus the luminosity at 5100\,\AA\ after
	 correction for the contribution from host galaxy starlight.}
\end{figure}

\clearpage

\end{document}